# NQR frequencies of anhydrous carbamazepine polymorphic phases

Bonin C. J.[a,*] , Amar A.[a] and Pusiol D. J.[a,b]

a Facultad de Matemática, Astronomía y Física, Universidad Nacional de Córdoba, Ciudad Universitaria, 5000 Córdoba, Argentina, Tel: ++54 351 4334051 - Fax: ++54 351 4334054.

b SpinLock S.R.L., C. de Arenal 1020, X5000GZU Córdoba, Argentina. www.spinlock.com.ar

* Corresponding author cbonin@famaf.unc.edu.ar

ABSTRACT

In this work we propose the Nuclear Quadrupole Resonance (NQR) technique as an analytical method suitable for polymorphism detection in active parts (or active principles) of pharmaceuticals with high pharmacological risk. Samples of powder carbamazepine (5H-dibenz(b,f)-azepine-5-carboxamide) are studied. In its anhydrous state, this compound presents at least three different polymorphic forms: form III, the commercial one, form II, and form I. Of these, only form III possesses desirable therapeutic effects. By using the NQR technique, it was possible to characterize two of the three polymorphic phases (I and III) for anhydrous carbamazepine in few minutes at room temperature, detecting the characteristic frequencies of $^{14}N$ nuclei (I=1) present in their chemical composition and in the frequency range 2.820—3.935 MHz. For form II, characteristic lines were not detected within this range of frequencies. The lines detected for form III are centered at the frequencies $\nu_1^{III}$=3.876 MHz, $\nu_2^{III}$=3.829 MHz, and $\nu_3^{III}$=2.909 MHz; and for form I centered at $\nu_1^{I}$=3.876 MHz, $\nu_2^{I}$=3.840 MHz, $\nu_3^{I}$=2.911 MHz, and $\nu_4^{I}$=2.830 MHz.

## I. INTRODUCTION

In the last decades, it has been proved that once a compound with pharmacological applications is developed, the characterization of its solid state is one of the most important stages for the development of the pharmaceuticals. An appropriate characterization is needed since solid pharmacological forms can be presented as different polymorphs, solvates or amorphous forms, which generally have different bioavailabilities and clinical use [1].

Polymorphism is the crystallization of the same compound in more than one distinct crystal architecture and is associated with different crystal packing arrangements. This phenomenon is very common in pharmacological drugs, as it is the case of carbamazepine (5H-dibenz(b,f)-azepine-5-carboxamide). This drug whose chemical formula is $C_{15}H_{12}N_2O$ has been in routine use in the treatment of epilepsy and trigeminal neuralgia for over 30 years. Three different polymorphic forms have been confirmed for the anhydrous carbamazepine [2]: form III, the commercial one, form I, and form II (nomenclature as Krahn and Mielck [3] and Behme and Brooke [4]); nevertheless, certain analytical results suggest the existence of additional forms and the formation of mixed crystals.

In spite of possessing the same active part (molecule), polymorphs have different chemical and physical properties; they have different melting points, different chemical reactivity, different dissolution rates, and different bioavailability (polymorph III of carbamazepine is the only one that has desirable therapeutic effects). Polymorphs can be interconverted by phase transformations or by a solvent-mediated process. Phase transformations can also be induced by heat or mechanical stress [2].

It is important to highlight that the Food and Drug Administration requires analytical procedures for the detection of polymorphic, hydrated, or amorphous forms in the drug substances, and that in certain cases the polymorphs of an active part pass the conventional pharmacological tests, since these analyze the compound mainly in solutions. For this reason, it is essential to have other techniques implemented to study these phenomena.

Since the sixties, it is well known that the Nuclear Quadrupole Resonance (NQR) [5, 6] technique is useful for studying basic aspects of crystalline phase transitions that involve different polymorphs. This technique together with the NMR (Nuclear Magnetic Resonance) has proved to be very sensitive for the microscopic study of diverse dynamic and structural properties of the atoms that constitute the solids. In particular, the sensitivity of the NQR frequency in the face of structural changes in the crystal converts this into a very appropriate spectroscopic technique for the detection of these structural changes in organic compounds, as well as for the comparative study of the molecular dynamics corresponding to different polymorphic forms, polymorphic phases transitions, etc. (see Refs. 5--10 and references therein). Therefore, the NQR technique is a valid alternative to compare active parts, avoiding the need of using bioavailabilities studies *in vivo*.

With regard to X-Ray powder diffraction and calorimetry techniques, the NQR implementation is technologically simpler and cheaper. In relation to the first one, it is not necessary to irradiate the sample (and eventually the environment) with ionizing radiations, but short and innocuous long wavelength radio frequencies (rf) pulses. As regards the second one, the analysis is performed at room temperature and in few seconds, without the necessity of high temperatures that in many cases lead to the decomposition of the compound. Another important aspect to be pointed out is that samples do not require previous preparation and can be used again. On the other hand, the analysis can be carried out directly in the pure drug as in the medication and it can be quantitative, regarding the total of the active part contained in the pharmaceutical formulation. The time required for the analysis can be reduced to a few seconds, and in some favorable cases, to even shorter periods. Since the action on the sample is innocuous, it is feasible to develop devices to control the final product directly in the production line. The analytic equipment can be implemented daily and controlled by means of a PC, so that highly qualified staff is not needed for its operation.

The aim of this work is to characterize the different polymorphic states of anhydrous carbamazepine (forms I, II and III) by the $^{14}N$ (I=1) NQR, through the evaluation of the characteristic NQR frequencies at room temperature.

## II. EXPERIMENTAL

The experiments were carried out using $^{14}N$ nuclei in a powder sample of carbamazepine, at room temperature, and the SpinLock QR20 broad band (0.4–20MHz) NQR spectrometer (see www.spinlock.com.ar for technical details). The signals were processed with an *had hoc* software provided by the SPINLOCK Software Department.

### A. MATERIALS

Carbamazepine samples in their polymorphic phases I, II, and III (nomenclature as Refs. 3 and 4) were provided by the Pharmaceutical Laboratory QUÍMICA LUAR S.R.L. (Córdoba-Argentina). At first sight, these polymorphs in their pure state look like a white thick powder for polymorphs I and III, and a white fine powder for polymorph II. Carbamazepine form III was analyzed by X-Ray Powder Diffraction and Fourier Transform Infrared Spectroscopy techniques according to USP XXIV in the National Organism CEPROCOR (Córdoba-Argentina). The analysis determined that the sample corresponds to carbamazepine polymorph III (monoclinic structure [11]) without the presence of other polymorphs. Technical characteristics of carbamazepine can be found in Refs. 2—4, 11—15.

B. SAMPLES PREPARATION

The samples were vacuum-packed in glass vessels of 113 $cm^3$. Two of them were filled with 50g carbamazepine polymorph I and III, respectively; while the other was filled with 43g of the form II. According to the literature [13], polymorph I (trigonal form) was obtained by heating form III in an oven at 140ºC for 9 h; polymorph II was obtained from freshly prepared carbamazepine dihydrate by dehydrating at 20ºC in a vacuum desiccator over $P_2O_5$ Ref. 3; and form III was obtained by cooling boiled acetone solution of commercial carbamazepine [13].

C. EXPERIMENTAL TECHNIQUE

To find the NQR characteristic frequencies of $^{14}N$ [*] nuclei in polymorphs I, II, and III, a sweeping-frequency was made by 5kHz-steps over the 2.820--3.935 MHz range and retuning the probe each 40kHz. For spectral line detection, CPMG rf pulses sequence[16] was used, which consists of an initial $90^o|_x$ pulse, followed by a train of $180^o|_y$ pulses, i.e., 90º phase shift. The length for 90º and 180º rf pulses was about 35 and 50μs respectively. The interval between the $90^o|_x$ and $180^o|_y$ pulses was 800μs, and the interval between $180^o|_y$ pulses was 1600 μs. The length of the pulse train was 30ms. By this sequence, the echo signals, formed between the $180^o|_y$ pulses of the train, were acquired and added. In order to improve signal-to-noise ratio (S/N), the NQR echo signal is take as an average value of about 50 pulse train measurements. The spectra were obtained by Fast Fourier Transform (FFT) of the signals. It is important to wait for some time after each pulse train so that the $^{14}N$ spin system relaxes and reaches thermal equilibrium with surroundings (lattice).

[*] Generally, the NQR characteristic frequencies of $^{14}N$ nuclei are in the frequency band 2—5 MHz.

Generally, a $5T_1$-order delay is chosen where $T_1$ is the spin-lattice relaxation time [16, 17]. In order to minimize the measurement time, a 5s-delay was chosen.†

## III. RESULTS AND ANALYSIS

For polymorph III, three NQR characteristic narrow lines were detected, each belonging to the 5kHz order, likely to be compound and centered at the characteristic frequencies: $\nu_1^{III}=3.876\ MHz$, $\nu_2^{III}=3.829\ MHz$ and $\nu_3^{III}=2.909\ MHz$, respectively. The experimental spectrum for polymorph III is shown in Fig. 1.

The most intense line corresponds to frequency $\nu_1^{III}$ and the less intense one is centered at $\nu_3^{III}$ (see Fig. 1).

A more detailed analysis of spectral lines shows, in fact, that these are multiple Gaussians-like lines. The number of these composing lines, their intensities and their widths were determined by an appropriate software. For this analysis, the irrelevant points of lines were eliminated, taking into account only those whose intensities were greater than the average intensity of thermal-noise signal. The adopted criteria for composing lines number determination was the following: in the first place, the line was fit with a Gaussian curve of the form $y(x) = \left(a_0 / \sqrt{2\pi a_2}\right) \mathbf{exp}\left( -(x-a_1)^2 / \left(\sqrt{2} a_2\right)^2 \right)$ and then we continue adding other Gaussian curves, keeping $a_0$, $a_1$ and $a_2$ parameters free, until being able to reproduce the measured line by using the smallest number of such curves. Once the confidence parameters provided by the software were the optimum ones (correlation parameter $r^2 \sim 1$ and merit factor $F >> 1$) lines (or curves) were no longer added. On the other hand, we used other kinds of curves, namely Lorentzian-like and combinations of Lorentzians and Gaussians, and the best adjustments were obtained by Gaussian curves. The lines determined in this way are listed in Tables I(a)--I(c) (see Figs. 2(a)--2(c)).

† With the rf pulse sequence $90^o|_y$—$t_1$--$90^o|_y$--$\tau$--$90^o|_x$--$2\tau$--$90^o|_x$…, it was possible to measure the spin-lattice relaxation time $T_1$; and for the polymorph III at characteristic frequency $\nu_2^{III}=3.829$ MHz it was $T_1=(3697\pm123)$ms.

In polymorph I, four NQR characteristic lines were detected, each belonging to the 10kHz order and also likely to be compound. These lines are centered at: $\nu_1^I=3.876\ MHz$, $\nu_2^I=3.840\ MHz$, $\nu_3^I=2.911\ MHz$ and $\nu_4^I=2.830\ MHz$, respectively, and they are shown in Fig. 3.

The most intense line corresponds to the frequency $\nu_1^I$, $\nu_2^I$ follows in intensity, and the less intense one is centered at $\nu_3^I$ (see Fig. 3).

For the determination of composing lines, we followed the same procedure as for polymorph III, obtaining the results shown in Tables II(a)—II(d) (see Figs. 4(a)—4(d)).

## IV. CONCLUSIONS

A complete evaluation of possible variations in crystallography of compounds of pharmacological interest is now essential for the development of new and more effective medicines. The sensitivity of NQR frequency in the face of structural changes in the crystal turns this into a very appropriate technique for polymorphisms characterization in active parts, avoiding the need of carrying out bioequivalence studies *in vivo*, which come to be extremely risky for individuals. It is quite feasible to develop devices to be used in the pharmaceutical and chemical industry since they are relatively economical and easy to be implemented.

The literature shows that the structure of polymorph I is more complex than that of polymorph III because the form I presents a greater disorder in its crystal structure. Polymorph I (trigonal structure) has a higher number of molecules per unit cell (eighteen), while polymorph III (monoclinic structure) has four molecules per unit cell[15]. These characteristics are proved by observing the corresponding spectra, which are experimentally measured and shown in Sec. III. For polymorph III, three NQR characteristic lines of $^{14}N$ (I=1) nuclei were detected, each belonging to the 5kHz order, likely to be compound and centered at the following frequencies: $\nu_1^{III}=3.876\ MHz$, $\nu_2^{III}=3.829\ MHz$, and $\nu_3^{III}=2.909\ MHz$ respectively. For polymorph I, on the other hand, four wider NQR lines were observed, each belonging to the 10kHz order and being likely to be compound, too. These lines are centered at the characteristic frequencies: $\nu_1^{I}=3.876\ MHz$, $\nu_2^{I}=3.840\ MHz$, $\nu_3^{I}=2.911\ MHz$, and $\nu_4^{I}=2.830\ MHz$ respectively. A greater number of spectral lines in polymorph I would be consistent with the presence of a higher number of non-equivalent $^{14}N$ nuclei in unit cell. As regards polymorph II, it was not possible to detect its characteristic frequencies within the frequencies range 2.820--3.935MHz. This might be due to the fact that this polymorph presents a disorder degree in its crystal structure that is even higher than those of polymorphs I and III.

## ACKNOWLEDGMENTS

The authors wish to thank to the participant institutions, to SPINLOCK S.R.L. (Córdoba-Argentina) for providing the NQR spectrometer and technical support and Pharmaceutical Laboratory QUÍMICA LUAR S.R.L. (Córdoba-Argentina) by the samples provided. C. J. B. thanks to Secretaría de Extensión Universitaria de la Universidad Nacional de Córdoba-Argentina for the financial support, to Lic. Germán D. Farrher for assistance with spectrometer and specially to Ms. María Laura Haye and Mrs. María Inés Fidalgo for checking the English of the manuscript.

Figure legends

Figure 1: NQR spectrum of carbamazepine polymorph III, measured at room temperature in frequency range 2.820—3.935 MHz.

Figure 2(a): Components of polymorph III line at $\nu_1^{III}$=3.876 MHz.

2(b): Components of polymorph III line at $\nu_2^{III}$=3.829 MHz.

2(c): Components of polymorph III line at $\nu_3^{III}$=2.909 MHz.

Figure 3: NQR spectrum of carbamazepine polymorph I, measured at room temperature in frequency range 2.820—3.935 MHz.

Figure 4(a): Components of polymorph I line at $\nu_1^{I}$=3.876 MHz.

4(b): Components of polymorph I line at $\nu_2^{I}$=3.840 MHz.

4(c): Components of polymorph I line at $\nu_3^{I}$=2.911 MHz.

4(d): Components of polymorph I line at $\nu_4^{I}$=2.830 MHz.

Table I(a): Amplitudes and widths of the composing lines of the line centered at $\nu_1^{III}$=3.876.

| *Frequency (MHz)* | *Component frequencies (MHz)* | *Amplitudes (arbitrary units)* | *Composing line widths (KHz)* |
|---|---|---|---|
| $\nu_1^{III}$=3.876 | 3.873 | 242 | 1.5 |
| | 3.874 | 138 | 1 |
| | 3.875 | 899 | 2 |
| | 3.877 | 1368 | 2 |
| | 3.878 | 262 | 1.5 |
| | 3.880 | 185 | 1.5 |

Table I(b): Amplitudes and widths of the composing lines of the line centered at $\nu_2^{III}$=3.829.

| *Frequency (MHz)* | *Component frequencies (MHz)* | *Amplitudes (arbitrary units)* | *Composing line widths (KHz)* |
|---|---|---|---|
| $\nu_2^{III}$=3.829 | 3.824 | 168 | 2 |
| | 3.826 | 60 | 1 |
| | 3.829 | 1141 | 3.5 |
| | 3.831 | 223 | 2.5 |
| | 3.834 | 102 | 1 |
| | 3.835 | 192 | 1.5 |

Table I(c): Amplitudes and widths of the composing lines of the line centered at $\nu_3^{III}$=2.909.

| ***Frequency (MHz)*** | ***Component frequencies (MHz)*** | ***Amplitudes (arbitrary units.)*** | ***Composing line widths (KHz)*** |
|---|---|---|---|
| $\nu_3^{III}$=2.909 | 2.903 | 25 | 3.5 |
| | 2.907 | 17 | 2 |
| | 2.908 | 61 | 3 |
| | 2.910 | 298 | 4.5 |
| | 2.914 | 25 | 4 |
| | 2.919 | 14 | 4 |

Table II(a): Amplitudes and widths of the composing lines of the line centered at $\nu_1^I$=3.876. This line is composed by five Gaussian lines mounted on a base-line.

| ***Frequency (MHz)*** | ***Component frequencies (MHz)*** | ***Amplitudes (arbitrary units)*** | ***Composing line widths (KHz)*** |
|---|---|---|---|
| $\nu_1^I$=3.876 | 3.868 | 120 | 3.5 |
| | 3.869 | 80 | 4.5 |
| | 3.875 | 277 | 4 |
| | 3.880 | 17 | 2.5 |
| | 3.883 | 20 | 4 |

Table II(b): Amplitudes and widths of the composing lines of the line centered at $\nu_2^I$=3.840.

| ***Frequency (MHz)*** | ***Component frequencies (MHz)*** | ***Amplitudes (arbitrary units)*** | ***Composing line widths (KHz)*** |
|---|---|---|---|
| $\nu_2^I$=3.840 | 3.830 | 11 | 2 |
| | 3.833 | 26 | 3.5 |
| | 3.836 | 19 | 2 |
| | 3.840 | 207 | 4.5 |
| | 3.843 | 13 | 2 |
| | 3.847 | 18 | 7 |

Table II(c): Amplitudes and widths of the composing lines of the line centered at $\nu_3^I$=2.911.

| ***Frequency (MHz)*** | ***Component frequencies (MHz)*** | ***Amplitudes (arbitrary units)*** | ***Composing line widths (KHz)*** |
|---|---|---|---|
| $\nu_3^I$=2.911 | 2.897 | 2 | 4 |
| | 2.901 | 4 | 3 |
| | 2.907 | 18 | 5 |
| | 2.912 | 39 | 8.5 |
| | 2.920 | 5 | 2.5 |
| | 2.921 | 2 | 1.5 |

Table II(d): Amplitudes and widths of the composing lines of the line centered at $\nu_4^I$=2.830.

| ***Frequency (MHz)*** | ***Component frequencies (MHz)*** | ***Amplitudes (arbitrary units)*** | ***Composing line widths (KHz)*** |
|---|---|---|---|
| $\nu_4^I$=2.830 | 2.819 | 7 | 5 |
| | 2.822 | 10 | 3 |
| | 2.826 | 16 | 4.5 |
| | 2.830 | 6 | 3 |
| | 2.831 | 39 | 6.5 |
| | 2.837 | 5 | 3 |
| | 2.840 | 3 | 3.5 |

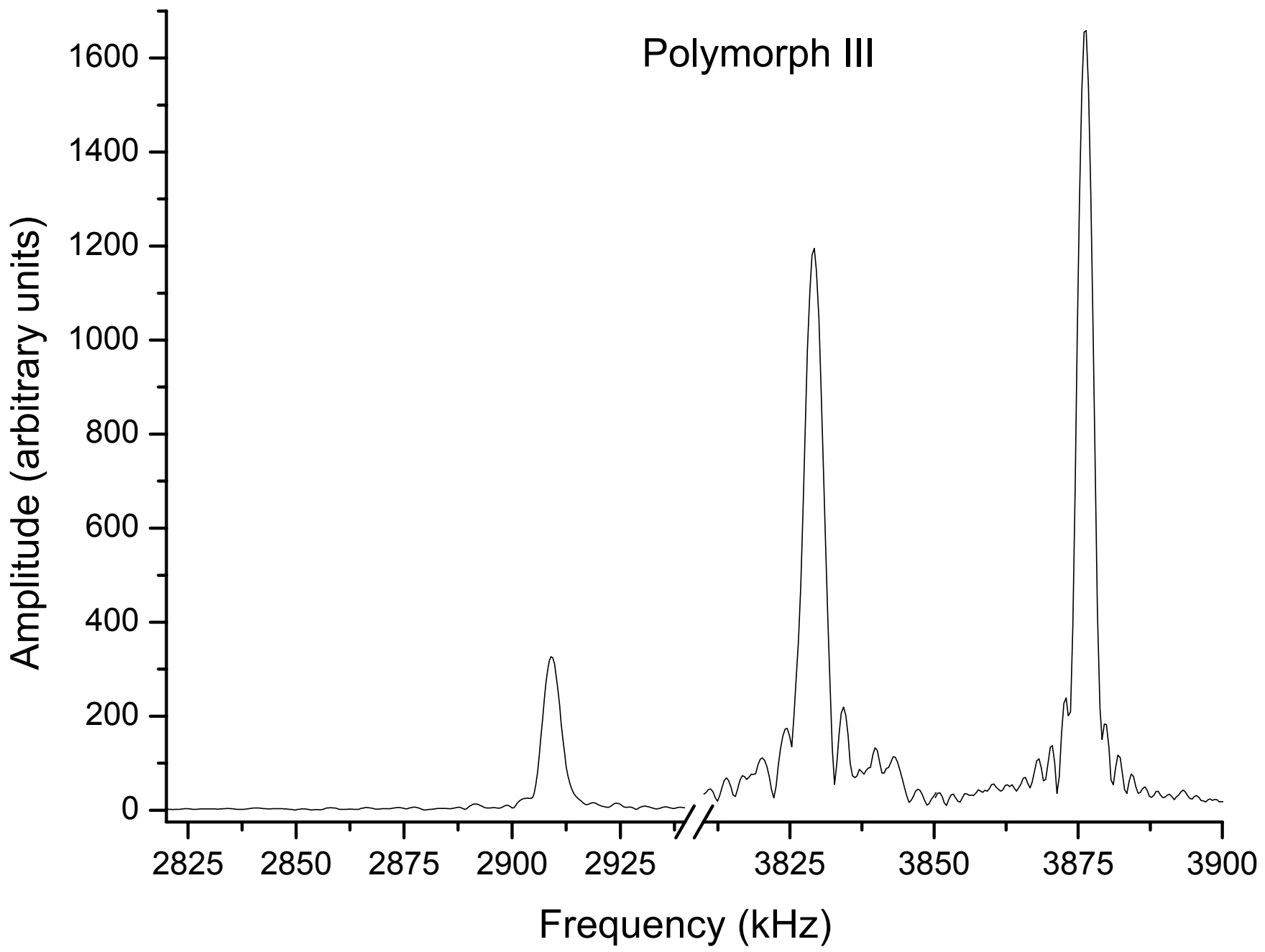

Figure 1

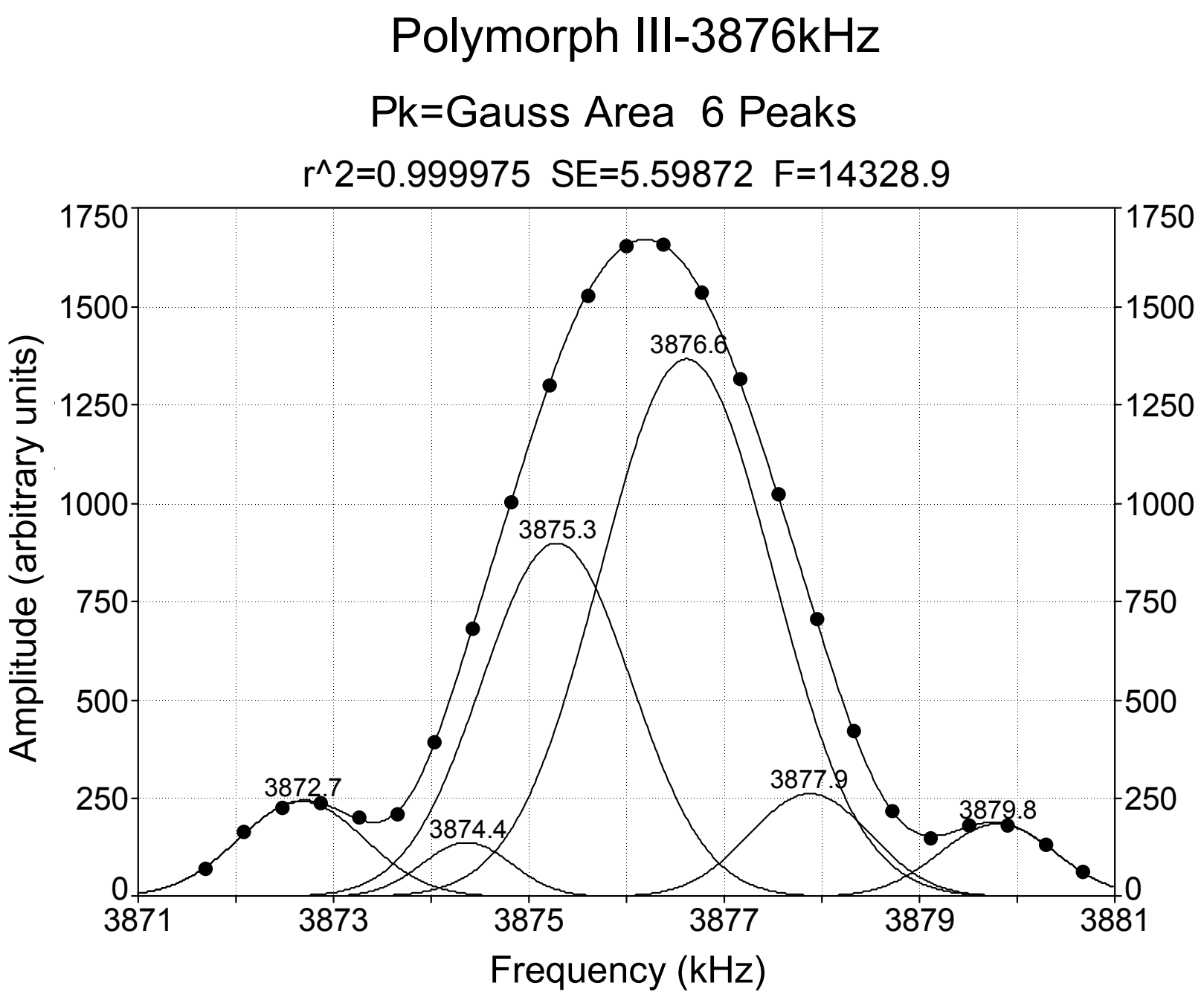

Figure 2(a)

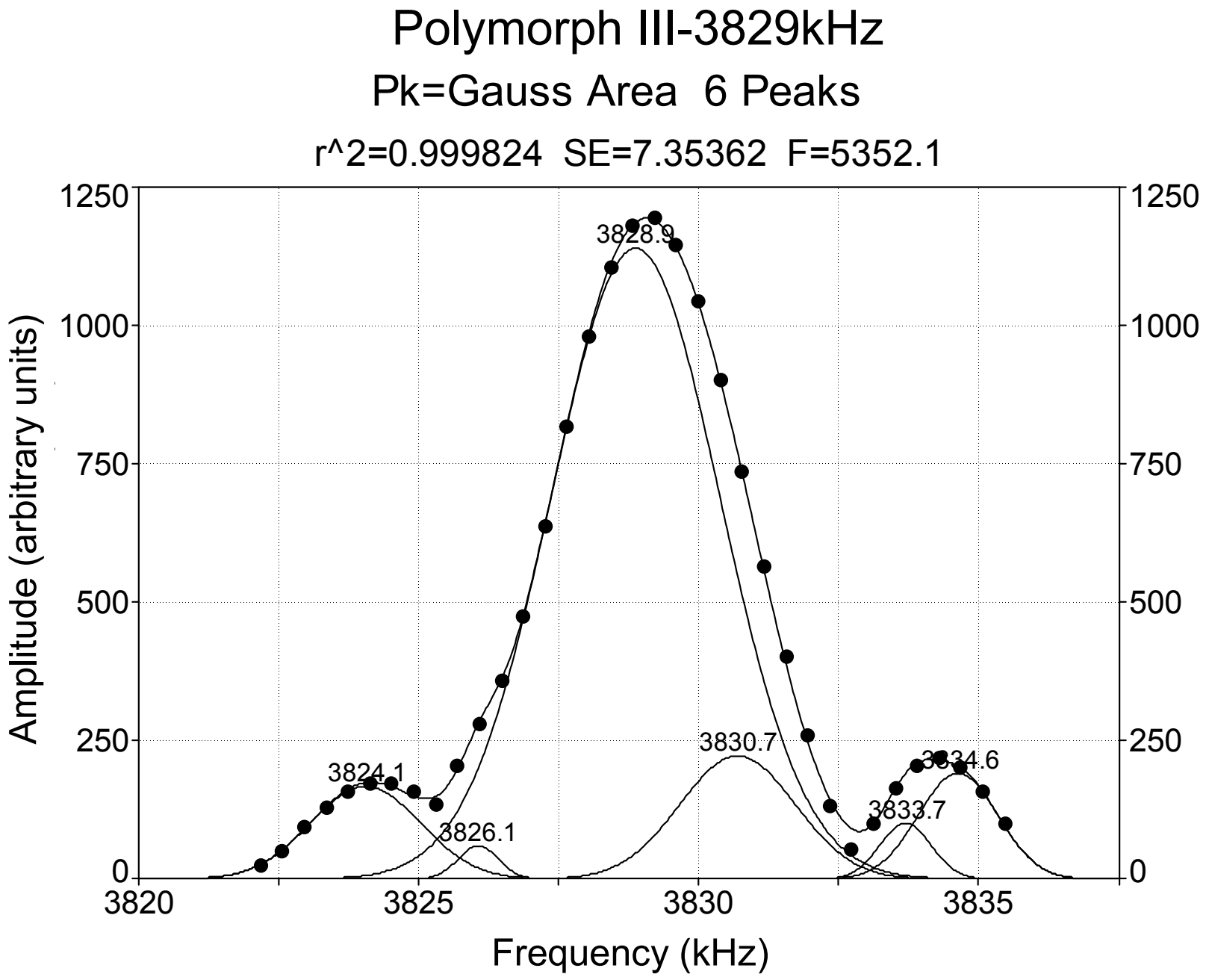

Figure 2(b)

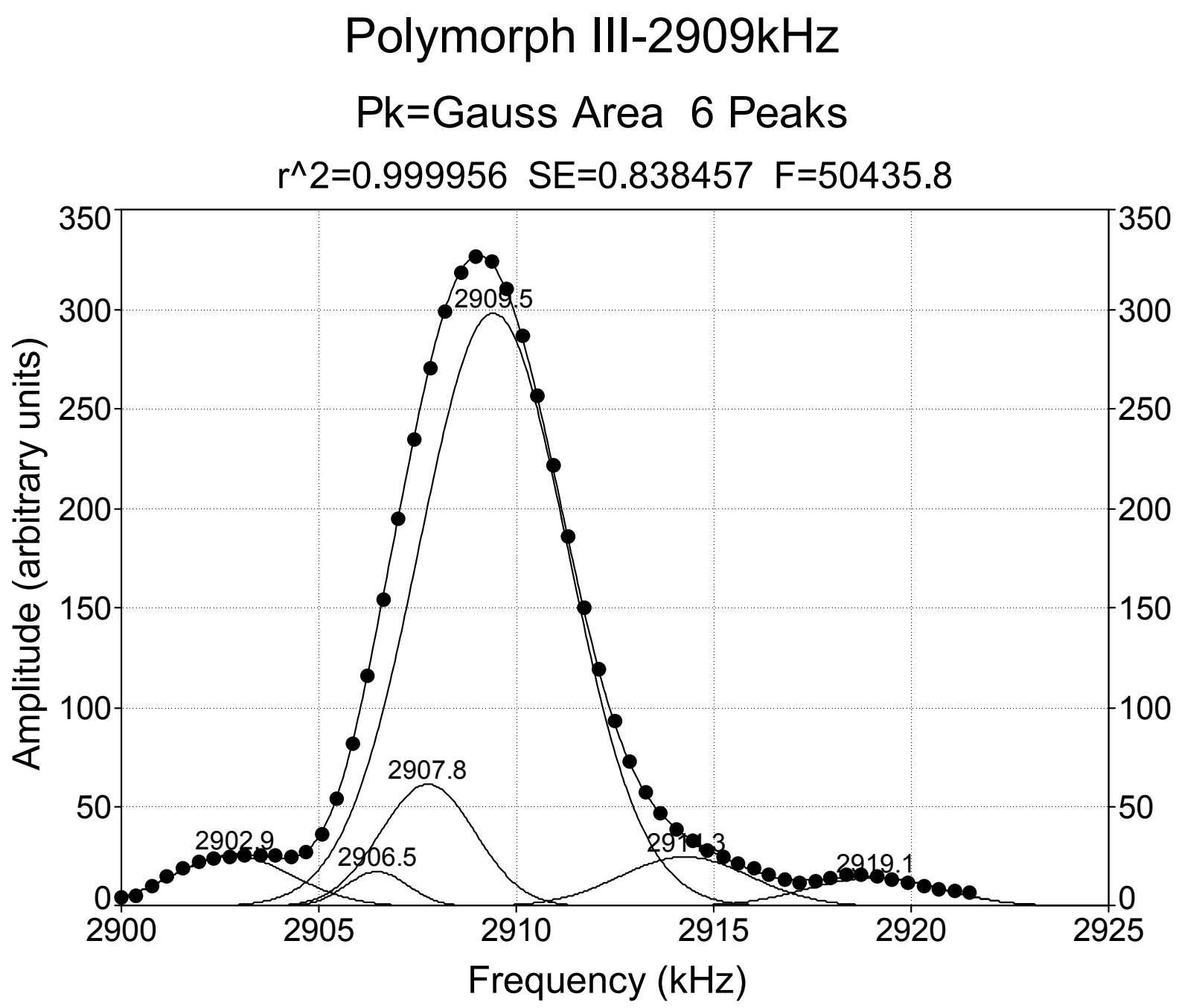

Figure 2(c)

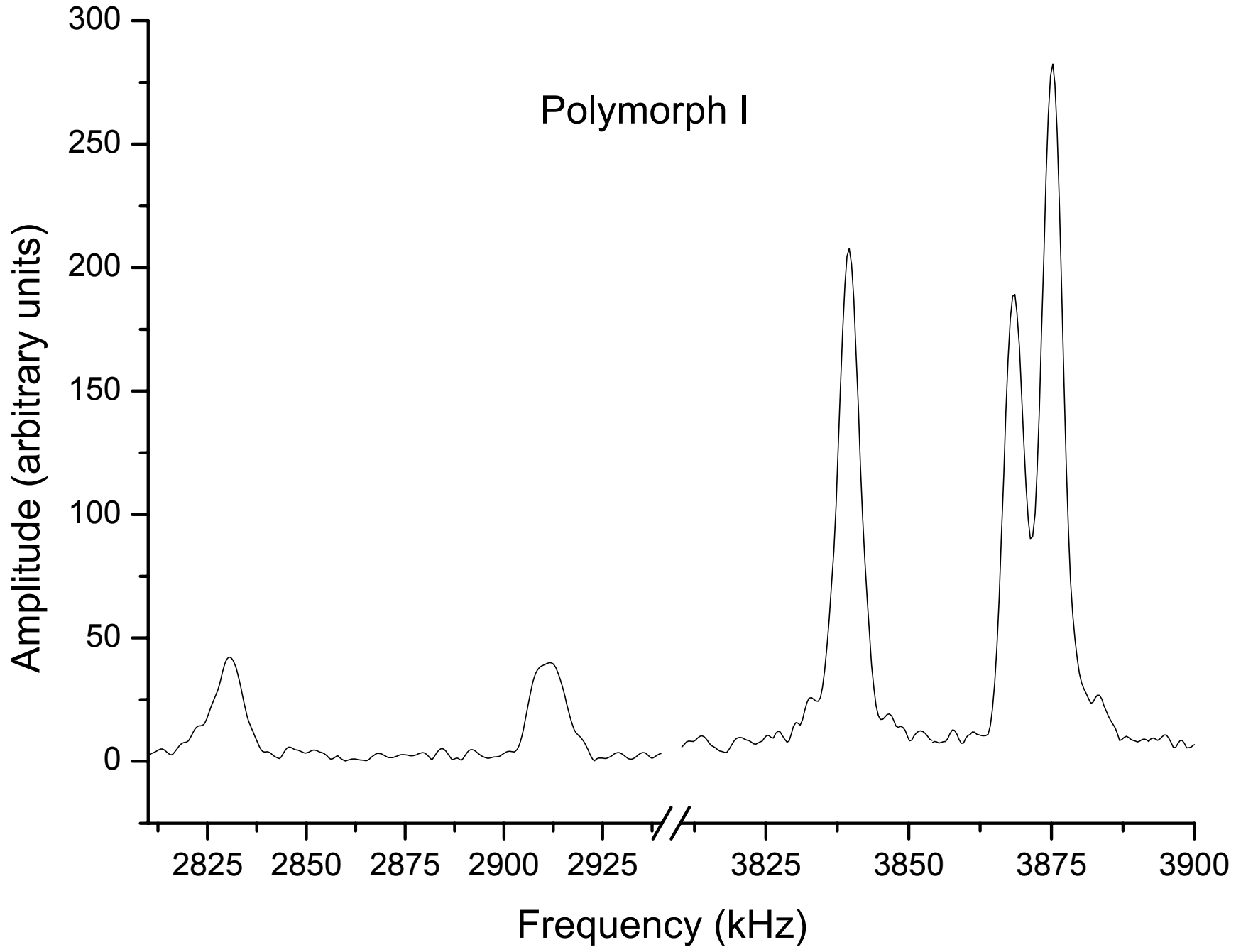

Figure 3

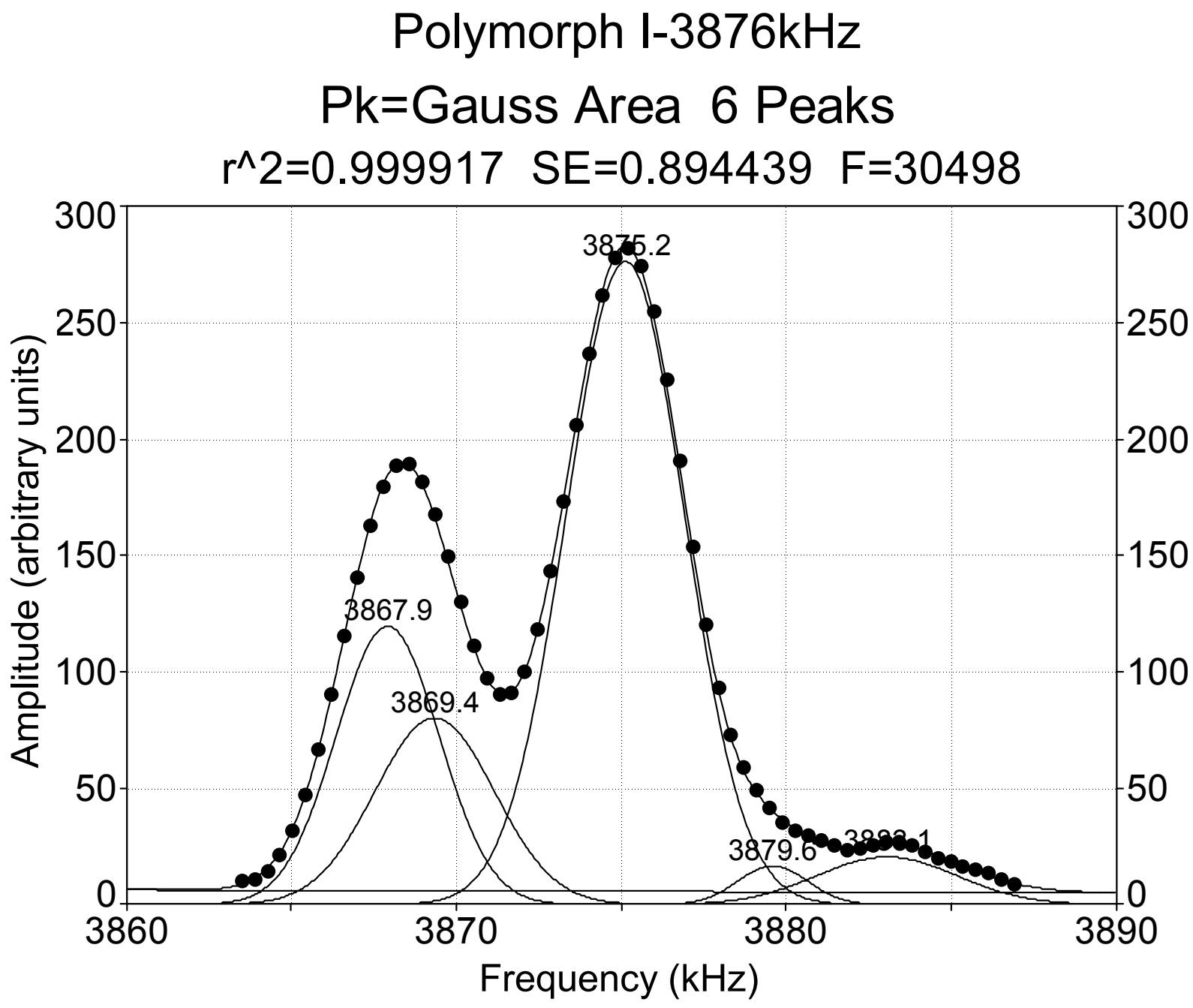

Figure 4(a)

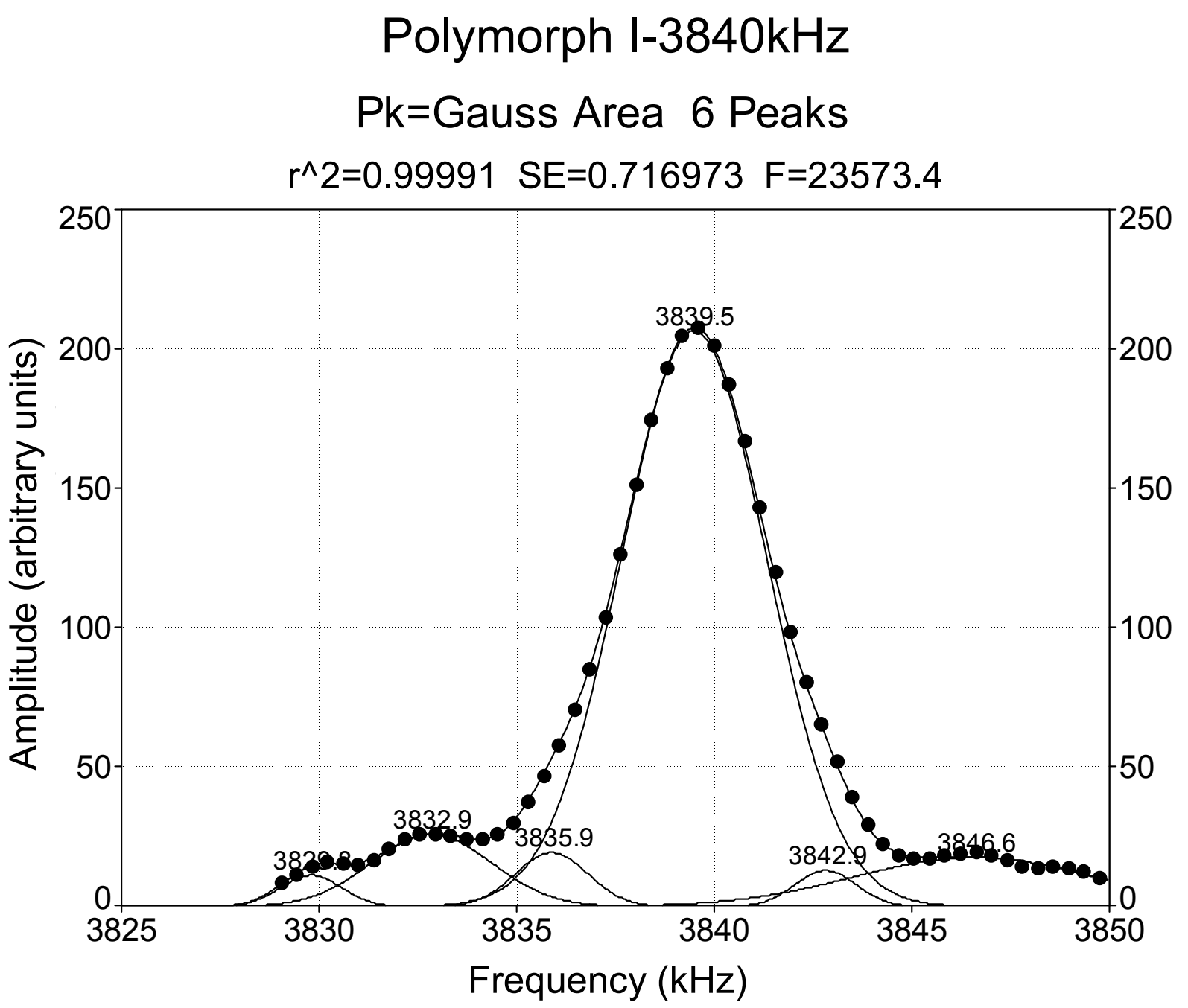

Figure 4(b)

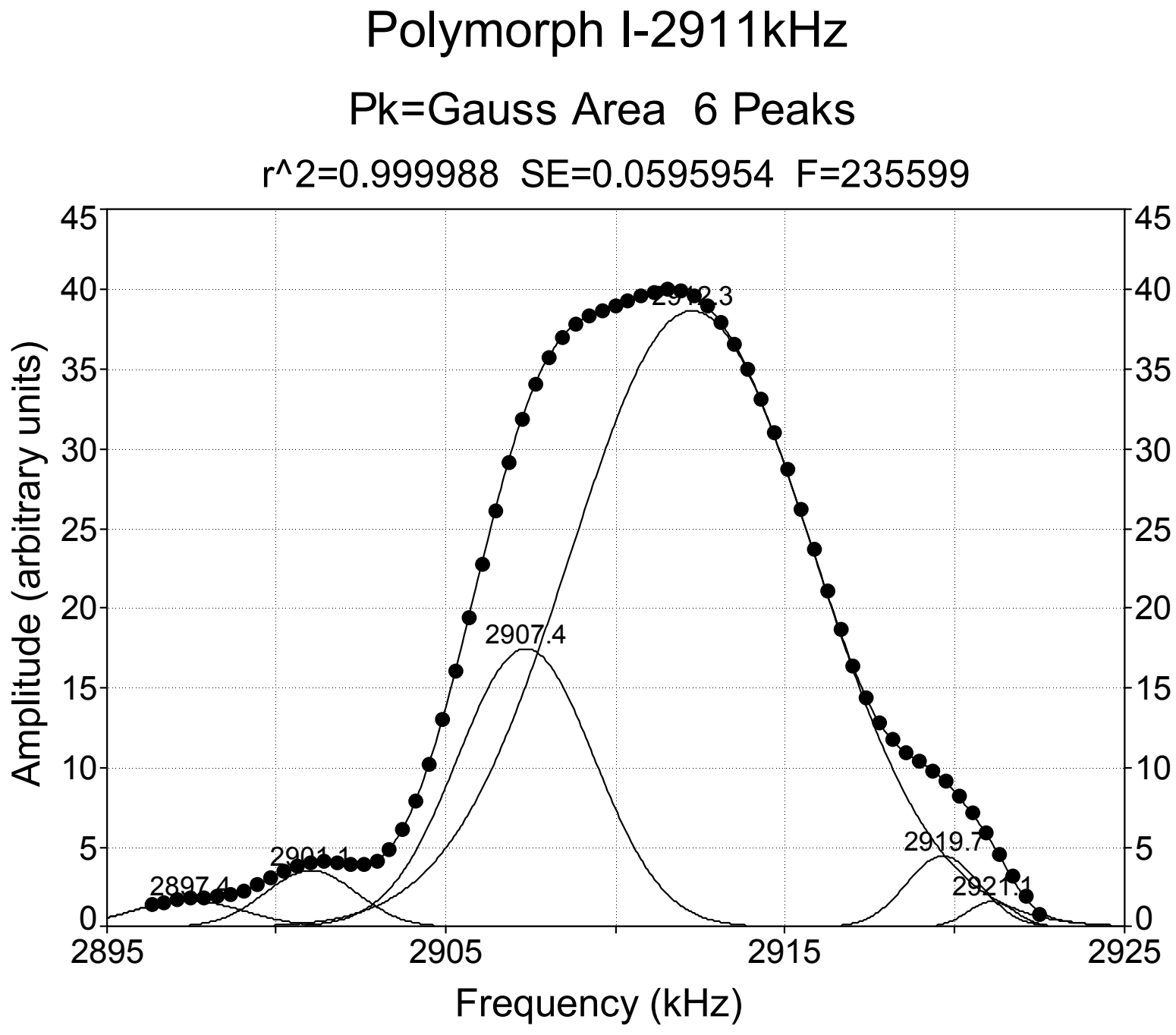

Figure 4(c)

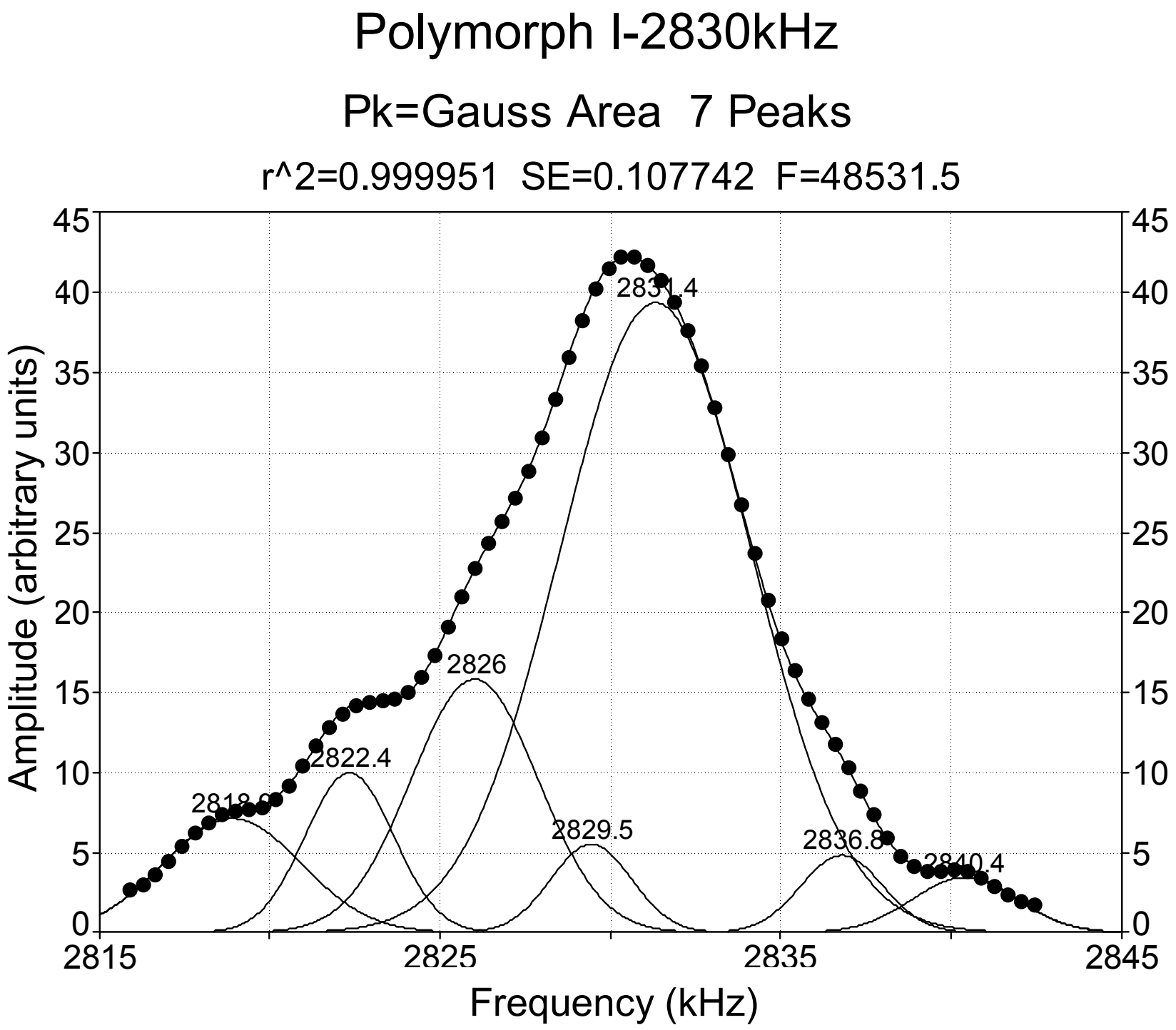

Figure 4(d)